\documentclass[pra]{revtex4-2}
 \usepackage{amssymb} \usepackage{graphicx}
\usepackage[utf8]{inputenc}
\usepackage[english]{babel} 
\usepackage{amsthm}
 \usepackage{amsmath}

\begin{document}
 \title{Extension of Poincare superalgebra via hidden symmetries}

\author{\"{U}mit Ertem}
\email{umitertem@ankara.edu.tr}
\affiliation{Department of Software Engineering, Ankara University,\\
Faculty of Engineering, 06830 G\"olba\c{s}{\i}, Ankara, T\"urkiye}

\begin{abstract}

Isometries of a manifold and special types of spinors called Killing spinors which are related to Killing vector fields constitute a symmetry superalgebra structure corresponding to Poincare superalgebra. The antisymmetric generalization of Killing vector fields to higher degree forms which are called Killing-Yano forms are considered. The odd part of the Poincare superalgebra is extended from the space of Killing vector fields and their rotational representation to the space of special odd Killing-Yano forms and special even closed conformal Killing-Yano forms. The even part is again corresponds to the space of Killing spinors. By defining the relevant brackets, we show that hidden symmetries and Killing spinors together define an extension of the Poincare superalgebra.

\end{abstract}

\maketitle

\section{Introduction}
Isometries of a manifold are generated by Killing vector fields which determine the symmetry structure of a manifold. Killing vector fields constitute a Lie algebra structure with respect to the Lie bracket of vector fields. On a spin manifold, besides Killing vectors, one can also define a special class of spinors called Killing spinors related to the symmetry structure of the manifold, namely the bilinear 1-forms that are constructed from Killing spinors correspond to Killing vector fields \cite{Lichnerowicz, Aleksy Cortes}. Killing vector fields and Killing spinors together constitute a symmetry superalgebra structure by defining symmetry operators of Killing spinors corresponding to spinorial Lie derivative \cite{Klinker, Medeiros Hollands}. Moreover, the space of Killing vector fields $V$, their rotational representation $so(V)$ and the space of Killing spinors $S$ define the Poincare superalgebra whose even part corresponds to $V\oplus so(V)$ and odd part corresponds to $S$ \cite{ACDV}. Besides defining the symmetry superalgebra structures on manifolds, Poincare superalgebra has also importance in the classification of supergravity backgrounds since they can be classified by filtered deformations of the subalgebras of the Poincare superalgebra \cite{OFarrill Santi1, OFarrill Santi2}. However, Killing vector fields have antisymmetric generalizations to higher degree forms called Killing-Yano (KY) forms which correspond to the hidden symmetries of the manifold \cite{Yano, Semmelmann, Kastor Ray Traschen, AEOV}. Higher degree bilinear forms of Killing spinors correspond to special KY and special closed conformal KY forms \cite{Acik Ertem21}. The extension of symmetry superalgebras to include hidden symmetries are proposed in the literature \cite{Acik Ertem21}, however the possible extension of the Poincare superalgebra to include hidden symmetries can also be investigated.

In this paper, we consider the structure of Poincare and other symmetry superalgebras on a manifold and define the possible extensions of the Poincare superalgebra to include higher degree hidden symmetries of the manifold. The Poincare superalgebra is written in terms of differential forms and rotational representation of Killing vector fields are defined in terms of Clifford bracket and contracted wedge products. We extend the structure of Poincare superalgebra by defining the generalization of the spinorial Lie derivative as the symmetry operators of Killing spinors and generalization of the rotational representation of higher degree KY forms in terms of contracted wedge products. We give the proofs of the extended Poincare superalgebra structure by investigating the brackets of the extended superalgebra. Possibility of generalizations of the classification of supergravity backgrounds by using these extended superalgebras are also investigated.

The paper is organized as follows. In Section II, the structure of the Poincare superalgebra is summarized. Section III includes the structure of symmetry superalgebras including hidden symmetries. In Section IV, the extensions of the Poincare superalgebra to include higher degree hidden symmetries are defined and related proofs are given. Section V concludes the paper.

\section{Poincare superalgebra}

Let us consider a manifold $M$. Generators of isometries on $M$ are defined as vector fields $K$ defined on the tangent bundle $TM$ which satisfy the following Killing equation
\[
\mathcal{L}_{K}g=0
\]
where $\mathcal{L}$ is the Lie derivative and $g$ is the metric tensor defined on $M$. This equality can also be written in terms of the Levi-Civita connection $\nabla$ and the metric dual 1-form $\widetilde{K}$ as
\[
\nabla_X\widetilde{K}=\frac{1}{2}i_Xd\widetilde{K}
\]
where $X$ is any vector field, $i_X$ is the contraction or interior product with respect to $X$ and $d$ is the exterior derivative. Moreover, on a spin manifold, one can also define spinor fields and the spinors $\psi$ that satisfy the following equation are called Killing spinors
\[
\nabla_X\psi=\lambda\widetilde{X}.\psi.
\]
where $\lambda$ is a constant and $.$ denotes the Clifford product. By using Killing vectors and Killing spinors one can define a Poincare superalgebra structure on $M$. Killing vectors correspond to the isometries of the background manifold and they constitute a Poincare Lie algebra structure $V\oplus so(V)$ where the translational part $V$ corresponds to the space of Killing vectors and the rotational part $so(V)$ corresponds to the rotational representation of Killing vectors. Killing vectors themselves satisfy a Lie algebra structure and the rotational representation of a Killing vector $K$ is defined by the Clifford bracket action of the exterior derivative of the metric dual of $K$ which is denoted by $\widetilde{K}$ and the action is written as
\[
[d\widetilde{K}, \, ]_{Cl}=-2d\widetilde{K}\underset{1}{\wedge}
\]
where we have used the fact that the Clifford bracket including a 2-form can be written as a contracted wedge product. This representation can be derived from a special rotation operator. We can construct a an operator corresponding to the difference of Lie and covariant derivatives with respect to a vector field $V$ as
\[
A_V=\mathcal{L}_V-\nabla_V
\]
and it can be shown that the rotational representation of a Killing vector $K$ can be found by using this difference operator. In fact the Lie derivative with respect to a Killing vector corresponds to a symmetry operator for Killing spinors and its action on a spinor can be written as
\[
\mathcal{L}_K\psi=\nabla_K\psi+\frac{1}{4}d\widetilde{K}.\psi
\]
and its generalization to the action on Clifford forms $\phi$ can be stated as
\[
\mathcal{L}_K\phi=\nabla_K\phi+\frac{1}{4}[d\widetilde{K}, \phi]_{Cl}.
\]
So, the action of the difference operator on a Clifford form $\phi$ can be written as
\begin{eqnarray}
A_K\phi&=&\mathcal{L}_K\phi-\nabla_K\phi\nonumber\\
&=&\frac{1}{4}[d\widetilde{K}, \phi]_{Cl}
\end{eqnarray}
\\
and if we restrict this action to any $q$-form component $\phi_q$, we have
\begin{eqnarray}
(A_K\phi)_q&=&\frac{1}{4}([d\widetilde{K}, \phi])_q\nonumber\\
&=&-\frac{1}{2}d\widetilde{K}\underset{1}{\wedge}\phi_q.
\end{eqnarray}
where we have used the definition of the Clifford commutator in terms of the contracted wedge products. Then, the difference operator corresponds to the rotational representation of Killing vectors acting on a $q$-form. Namely, it rotates the $q$-form $\phi_q$ to another $q$-form $(A_K\phi)_q$ and the rotation operator is $-\frac{1}{2}d\widetilde{K}\underset{1}{\wedge}$.

So, the brackets of the Poincare algebra are defined as follows. Lie bracket of vector fields is defined for the elements of $V$ or if the duals of Killing vectors are used then the Schouten-Nijenhuis (SN) bracket, which is the generalization of the Lie bracket of vector fields to the higher degree differential forms, can also be used 
\[
[\widetilde{K}_1,\widetilde{K}_2]_{SN}=i_{X_a}\widetilde{K}_1\wedge\nabla_{X^a}\widetilde{K}_2-i_{X_a}\widetilde{K}_2\wedge\nabla_{X^a}\widetilde{K}_1
\]
For the bracket between the elements of $V$ and $so(V)$, we have
\[
[d\widetilde{K}_1, \widetilde{K}_2]_{Cl}=-2d\widetilde{K}_1\underset{1}{\wedge}\widetilde{K}_2
\]
and for the bracket of the elements of $so(V)$, we have
\[
[d\widetilde{K}_1, d\widetilde{K}_2]_{Cl}=-2d\widetilde{K}_1\underset{1}{\wedge}d\widetilde{K}_2
\]
By using the definition of the SN bracket, we can calculate the exterior derivative of SN bracket of two Killing 1-forms as
\begin{eqnarray}
d[\widetilde{K}_1, \widetilde{K}_2]_{SN}&=&-\frac{1}{2}d\widetilde{K}_1\underset{1}{\wedge}d\widetilde{K}_2+2R_{bc}\wedge i_{X^c}\widetilde{K}_1\wedge i_{X^b}\widetilde{K}_2\nonumber\\
&=&\frac{1}{4}[d\widetilde K_1, d\widetilde{K}_2]_{Cl}-4R(K_1, K_2).
\end{eqnarray}
where $R(X,Y)$ denotes the curvature operator and $R_{ab}$ are curvature 2-forms. So, we can also write the bracket of the $so(V)$ part in terms of the SN bracket as
\[
[d\widetilde{K}_1, d\widetilde{K}_2]_{Cl}=4d[\widetilde{K}_1, \widetilde{K}_2]_{SN}+16 R(K_1, K_2).
\]
The Poincare Lie algebra structure can be extended to a Lie superalgebra structure called Poincare superalgebra. If we denote the space of Killing spinors as $S$, the Poincare superalgebra is defined as $p_0\oplus p_1=V\oplus so(V)\oplus S$ where the even part is the Poincare algebra $p_0=V\oplus so(V)$ and odd part is $p_1=S$. The brackets of the Poincare superalgebra is defined as follows. For $K \in V$ and $\psi, \psi_1, \psi_2 \in S$, the even-even bracket $p_0\times p_0\rightarrow p_0$ is the bracket defined for the Poincare algebra. The even-odd bracket corresponds to the action of the spinorial Lie derivative on Killing spinors with respect to Killing vectors
\[
p_0\times p_1\rightarrow p_1
\]
\[
[(K, d\widetilde{K}), \psi]=\mathcal{L}_K \psi=\nabla_K\psi+\frac{1}{4}d\widetilde{K}.\psi
\]
The odd-odd bracket is the bilinear forms of Killing spinors defined as
\[
p_1\times p_1\rightarrow p_0
\]
\[
[\psi_1, \psi_2]=(\widetilde{(\psi_1\overline{\psi}_2)_1}, d\widetilde{(\psi_1\overline{\psi}_2)_1})
\] since bilinear 1-forms of Killing spinors correspond to the metric duals of Killing vectors. All of these brackets satisfy the Jacobi identities and the Poincare superalgebra has a Lie superalgebra structure \cite{ACDV}.

\section{Symmetry superalgebras of hidden symmetries}

Killing vector fields have antisymmetric generalizations to higher-degree differential forms.
If a $p$-form $\omega$ satisfies
\begin{equation}
\nabla_X\omega=\frac{1}{p+1}i_Xd\omega
\end{equation}
for any vector field $X$, then $\omega$ is called as a Killing-Yano (KY) $p$-form.
KY forms are special cases of conformal Killing-Yano (CKY) forms.
If a $p$-form $\omega$ satisfies
\begin{equation}
\nabla_X\omega=\frac{1}{p+1}i_Xd\omega-\frac{1}{n-p+1}\widetilde{X}\wedge\delta\omega
\end{equation}
for any vector field $X$, its metric dual $\widetilde{X}$ and the co-derivative operator $\delta$, then $\omega$ is called as a CKY $p$-form.
As can be seen from the last definition that the co-closed CKY forms satisfying $\delta\omega=0$ correspond to KY forms. On the other hand, another subset of CKY forms consists of closed CKY (CCKY) forms satisfying $d\omega=0$, namely the following equation
\begin{equation}
\nabla_X\omega=-\frac{1}{n-p+1}\widetilde{X}\wedge\delta\omega.
\end{equation}
CKY equation (5) has Hodge duality invariance, namely if $\omega$ is a CKY $p$-form then $*\omega$ is also a CKY $(n-p)$-form with $*$ denoting the Hodge map \cite{Semmelmann}. One can show that this property leads to the fact that for a KY $p$-form $\omega$, its Hodge dual $*\omega$ is a CCKY $(n-p)$-form and conversely for a CCKY $p$-form $\omega$, its Hodge dual $*\omega$ is a KY $(n-p)$-form. So, KY $p$-forms and CCKY $(n-p)$-forms are Hodge duals of each other. The integrability condition of the KY equation (4) can be found as follows \cite{Semmelmann, AEOV}
\begin{equation}
\nabla_{X_a}d\omega=-\frac{p+1}{p}R_{ab}\wedge i_{X^b}\omega
\end{equation}
and similarly for the CCKY equation (6), the integrability condition is \cite{Ertem16}
\begin{equation} 
\nabla_{X_a}\delta\omega=-\frac{n-p+1}{n-p}i_{X^c}\left(i_{X_b}R_{ca}\wedge i_{X^b}\omega\right).
\end{equation}
We can define subsets of KY and CCKY forms which satisfy special types of integrability conditions.
A KY $p$-form $\omega$ is called a special KY $p$-form, if it satisfies the following condition
\begin{equation}
\nabla_X d\omega=-c(p+1)\widetilde{X}\wedge\omega
\end{equation}
where $c$ is a constant. Similarly, a CCKY $p$-form is called a special CCKY $p$-form if it satisfies
\begin{equation}
\nabla_{X}\delta\omega=c(n-p+1)i_X\omega
\end{equation}
where $c$ is a constant.
Note that, in constant curvature manifolds with curvature 2-forms $R_{ab}=ce_a\wedge e_b$ for a constant $c$, the integrability conditions (7) and (8) reduces to (9) and (10), respectively. This means that in constant curvature manifolds, all KY forms are special KY forms and all CCKY forms are special CCKY forms. Obviously, this is not true in general.
On a manifold $M$, odd degree special KY forms satisfying (4) and (9) and even degree CCKY forms satisfying (6) and (10) constitute a Lie algebra structure with respect to the CKY bracket $[\,,\,]_{CKY}$ given as
\begin{equation}
[\omega_1, \omega_2]_{CKY}=\frac{(-1)^p}{n-q+1}\omega_1\wedge\delta\omega_2+\frac{1}{n-p+1}\delta\omega_1\wedge\omega_2
\end{equation}
in all dimensions \cite{Ertem16}. On the other hand, even degree special KY forms satisfying (4) and (9) and odd degree special CCKY forms satisfying (6) and (10) constitute a Lie algebra structure with respect to the Hodge star of the CKY bracket $*[\,,\,]_{CKY}$ in even dimensions.
On an $n$-dimensional manifold $M$ admitting Killing spinors, if the bilinears of Killing spinors correspond to the special odd KY and special even CCKY forms, then the space of Killing spinors and special odd KY and special even CCKY forms generated by them constitute a superalgebra structure which we denote by $\mathfrak{K}=\mathfrak{K}_0\oplus\mathfrak{K}_1$. The even part $\mathfrak{K}_0$ of the superalgebra is the Lie algebra of the special odd KY and special even CCKY forms with respect to the CKY bracket and the odd part $\mathfrak{K}_1$ is the space of Killing spinors. This superalgebra is called the generalized symmetry superalgebra and the bilinear operation is defined as follows. The even-even bracket is the CKY bracket of the special odd KY and special even CCKY forms defined in (11)
\begin{equation}
[\,,\,]_{CKY}:\mathfrak{K}_0\otimes\mathfrak{K}_0\rightarrow\mathfrak{K}_0
\end{equation}
The even-odd bracket is the symmetry operators of Killing spinors generated from the special odd KY and special even CCKY forms
\begin{equation}
L:\mathfrak{K}_0\otimes\mathfrak{K}_1\rightarrow\mathfrak{K}_1
\end{equation}
The odd-odd bracket is the $p$-form bilinears of Killing spinors written as
\begin{equation}
(\,\,)_p:\mathfrak{K}_1\otimes\mathfrak{K}_1\rightarrow\mathfrak{K}_0
\end{equation}
These brackets constitute a superalgebra structure.\\

\section{Extended Poincare superalgebras}

We propose an extension of the Poincare superalgebra structure to include higher degree KY forms besides Killing vectors. The even part of the Poincare superalgebra was equivalent to $V\oplus so(V)$ which the direct sum of the space of Killing vectors and their rotational representation that is equivalent to the space of 2-forms constructed from Killing vectors. We can generalize this construction to any higher degree KY $p$-form $\omega$ as follows. Now, we extend $V$ to the space of special odd KY forms $\Lambda^{odd}V$ and extend $so(V)$ to the space of special even CCKY forms $\Lambda^{even}V$. The symmetry operators of Killing spinors can be constructed from KY forms as generalizations of the Lie derivative of spinors with respect to Killing vectors \cite{Ertem16}. For a KY $p$-form $\omega$, the symmetry operators can be written as
\[
L_{\omega}\psi=(i_{X^a}\omega).\nabla_{X_a}\psi+\frac{p}{2(p+1)}d\omega.\psi
\]
and it can be generalized to the action on Clifford forms as
\[
L_{\omega}\phi=(i_{X^a}\omega).\nabla_{X_a}\psi+\frac{p}{2(p+1)}[d\omega.\psi]_{Cl}
\]
and we can define the difference operator is defined in this case as
\[
A_{\omega}=L_{\omega}-\nabla_{\omega}
\]
where the definition of $\nabla_{\omega}$ is given by
\[
\nabla_{\omega}=(i_{X^a}\omega).\nabla_{X_a}.
\]
So, the action of the difference operator on a Clifford form $\phi$ is written as
\[
A_{\omega}\phi=\frac{p}{2(p+1)}[d\omega.\psi]_{Cl}
\]
and if we restrict this action to any $q$-form component $\phi_q$, we have
\begin{eqnarray}
(A_{\omega}\phi)_q&=&\frac{p}{2(p+1)}([d\omega, \phi])_q\nonumber\\
&=&-\frac{2}{(p+1)^2(p-1)!}d\omega\underset{\frac{p+1}{2}}{\wedge}\phi_q.
\end{eqnarray}
where we have used the wedge product expansion of the Clifford bracket. Thus, the rotational representation of special KY $p$-forms acting on a $q$-form is given in that way. It rotates the $q$-form $\phi_q$ to another $q$-form $(A_{\omega}\phi)_q$ and the rotation operator is
\[
-\frac{2}{(p-1)!(p+1)^2}d\omega\underset{\frac{p+1}{2}}{\wedge}
\]
and it can be shown that the exterior derivative $d\omega$ of a special odd KY $p$-form $\omega$ corresponds to a special even CCKY $(p+1)$-form. For a special KY $p$-form $\omega$, we can write the following KY equation and the integrability condition
\begin{eqnarray}
\nabla_{X_a}\omega&=&\frac{1}{p+1}i_{X_a}d\omega\\
\nabla_{X_a}d\omega&=&-c(p+1)e_a\wedge \omega.
\end{eqnarray}
If we apply $-i_{X^a}$ to the second equality (16), we find
\begin{eqnarray}
-i_{X^a}\nabla_{X_a}d\omega=\delta d\omega&=&c(p+1)i_{X^a}(e_a\wedge\omega)\nonumber\\
&=&c(p+1)(n-p)\omega.
\end{eqnarray}
If we replace (17) in (16), we can easily write
\[
\nabla_{X_a}d\omega=-\frac{1}{n-(p+1)+1}e_a\wedge\delta d\omega
\]
which is the defining equation for $d\omega$ to be a CCKY $(p+1)$-form. Moreover, by applying the covariant derivative $\nabla_{X_a}$ to (17), we obtain
\begin{eqnarray}
\nabla_{X_a}\delta d\omega&=&c(p+1)(n-p)\nabla_{X_a}\omega\nonumber\\
&=&c(n-p)i_{X_a}d\omega
\end{eqnarray}
which is the integrability condition for $d\omega$ to be a special CCKY $(p+1)$-form. Then, we show that if $\omega$ is a special odd KY $p$-form, then $d\omega$ is a special even CCKY $(p+1)$-form. Now, we can define the brackets of the extended Poincare algebra. In fact, the bracket of the extended Poincare algebra can be defined as the CKY bracket. For the elements of $\Lambda^{odd}M$ such as a special odd KY $p$-form $\omega_1$ and a special odd KY $q$-form $\omega _2$, we have
\[
[\omega_1, \omega_2]_{SN}=i_{X^a}\omega_1\wedge\nabla_{X_a}\omega_2+(-1)^{pq}i_{X^a}\omega_2\wedge\nabla_{X_a}\omega_1
\]
and the action of $\Lambda^{odd}M$ on $\Lambda^{even}M$ is defined as the $q$-form projection of the Clifford bracket
\[
([d\omega_1, \omega_2]_{Cl})_q=-\frac{4}{(p+1)!}d\omega_1\underset{\frac{p+1}{2}}{\wedge}\omega_2.
\]
The bracket for $\Lambda^{even}$ elements is written as the $(p+q)$-form projection of the Clifford bracket
\[
([d\omega_1, d\omega_2]_{Cl})_{p+q}=-\frac{(p+q)}{(p+1)(q+1)}d\omega_1\underset{1}{\wedge}d\omega_2
\]
From the exterior derivative of the SN bracket of $\omega_1$ and $\omega_2$ which can be found as
\[
d[\omega_1,\omega_2]_{SN}=-\frac{(p+q)}{(p+1)(q+1)}d\omega_1\underset{1}{\wedge}d\omega_2-\frac{(-1)^p(p+q)}{pq}R_{bc}\wedge i_{X^c}\omega_1\wedge i_{X^b}\omega_2
\]
one can write the bracket for $\Lambda^{even}M$ as follows
\[
([d\omega_1, d\omega_2]_{Cl})_{p+q}=d[\omega_1, \omega_2]_{SN}+\frac{(-1)^p(p+q)}{pq}R_{bc}\wedge i_{X^c}\omega_1\wedge i_{X^b}\omega_2.
\]

Now, we can extend the Poincare superalgebra by defining the even part as the extended Poincare algebra. The odd part of the extended superalgebra will still be given by the space of Killing spinors $S$. The extended superalgebra is $e_0\oplus e_1=\Lambda^{odd}V\oplus\Lambda^{even}V\oplus S$ and the brackets are defined as follows. The even-even bracket
\[
[\,,\,]:e_0\times e_0\rightarrow e_0
\] can be defined as the brackets of the extended Poincare algebra. The even-odd bracket
\[
[\,,\,]:e_0\times e_1\rightarrow e_1
\]
is defined as the symmetry operators of Killing spinors
\begin{eqnarray}
[(\omega, -\frac{2}{(p-1)!(p+1)^2}d\omega\underset{\frac{p+1}{2}}{\wedge}), \psi]&=&L_{\omega}\psi\nonumber\\
&=&(i_{X^a}\omega).\nabla_{X_a}\psi+\frac{p}{2(p+1)}d\omega.\psi\nonumber\\
&=&(-(-1)^p\lambda p\omega+\frac{p}{2(p+1)}d\omega).\psi
\end{eqnarray}
where we have used the Killing spinor equation. The odd-odd bracket
\[
[\,,\,]:e_1\times e_1\rightarrow e_1
\]
is defined as
\[
[\psi_1,\psi_2]=\bigg((\psi_1\overline{\psi_2})_{p},-\frac{2}{(p-1)!(p+1)^2}d(\psi_1\overline{\psi_2})_{p}\underset{\frac{p+1}{2}}{\wedge}\bigg)
\]
where for a Killing spinor $\psi$, the odd degree spinor bilinear form $(\psi_1\overline{\psi_2})_{p}$ corresponds to a special odd KY $p$-form.

In fact, we can also extend $V$ to the space of special odd CCKY forms $\Lambda^{odd}V$ and extend $so(V)$ to the space of special even KY forms $\Lambda^{even}V$. For a special odd CCKY $p$-form $\omega$, the corresponding rotational representation can also be found from the symmetry operators of Killing spinors written in terms of CCKY forms. We can generalize the construction for KY forms to any higher degree CCKY $p$-form $\omega$ as follows. The symmetry operators of Killing spinors can be constructed from CCKY forms as generalizations of the Lie derivative of spinors with respect to Killing vectors \cite{Acik Ertem21}. For a CCKY $p$-form $\omega$, the symmetry operators can be written as
\[
L_{\omega}\psi=(i_{X^a}\omega).\nabla_{X_a}\psi+\frac{p}{2(n-p+1)}(\delta\omega)^{\xi}.\psi
\]
and it can be generalized to the action on Clifford forms as
\[
L_{\omega}\phi=(i_{X^a}\omega).\nabla_{X_a}\psi+\frac{p}{2(n-p+1)}[(\delta\omega)^{\xi}.\psi]_{Cl}
\]
and we can define the difference operator in this case as
\[
A_{\omega}=L_{\omega}-\nabla_{\omega}
\]
where the definition of $\nabla_{\omega}$ is again given by
\[
\nabla_{\omega}=(i_{X^a}\omega.\nabla_{X_a}).
\]
So, the action of the difference operator on a Clifford form $\phi$ is written as
\[
A_{\omega}\phi=\frac{p}{2(n-p+1)}[(\delta\omega)^{\xi}.\psi]_{Cl}
\]
and if we restrict this action to any $q$-form component $\phi_q$, we have
\begin{eqnarray}
(A_{\omega}\phi)_q&=&\frac{p}{2(n-p+1)}([(\delta\omega)^{\xi}, \phi])_q\nonumber\\
&=&-\frac{2}{(n-p+1)^2(n-p)!}(\delta\omega)^{\xi}\underset{\frac{n-p+1}{2}}{\wedge}\phi_q.
\end{eqnarray}
Thus, the rotational representation of CCKY $p$-forms acting on a $q$-form is given in that way. It rotates the $q$-form $\phi_q$ to another $q$-form $(A_{\omega}\phi)_q$ and the rotation operator is
\[
-\frac{2}{(n-p+1)^2(n-p)!}(\delta\omega)^{\xi}\underset{\frac{n-p+1}{2}}{\wedge}.
\]
It can be shown that $\delta\omega$ corresponds to a special even KY $(p-1)$-form. For a special CCKY $p$-form $\omega$, we can write the following CCKY equation and the integrability condition
\begin{eqnarray}
\nabla_{X_a}\omega&=&-\frac{1}{n-p+1}e_a\wedge\delta\omega\\
\nabla_{X_a}\delta\omega&=&c(n-p+1)i_{X_a}\omega.
\end{eqnarray}
If we apply $e_a\wedge$ to the second equality (20), we find
\begin{eqnarray}
e_a\wedge\nabla_{X_a}\delta\omega=d\delta\omega=cp(n-p+1)\omega
\end{eqnarray}
If we replace (21) in (20), we can easily write
\[
\nabla_{X_a}\delta\omega=\frac{1}{p}i_{X_a}d\delta\omega
\]
which is the defining equation for $\delta\omega$ to be a KY $(p-1)$-form. Moreover, by applying the covariant derivative $\nabla_{X_a}$ to (21), we obtain
\begin{eqnarray}
\nabla_{X_a}d\delta\omega&=&cp(n-p+1)\nabla_{X_a}\omega\nonumber\\
&=&-cpe_a\wedge\delta\omega
\end{eqnarray}
which is the integrability condition for $\delta\omega$ to be a special KY $(p-1)$-form. Then, we show that if $\omega$ is a special odd CCKY $p$-form, then $\delta\omega$ is a special even KY $(p-1)$-form. In fact, the symmetry operators constructed from odd CCKY forms and algebra structure for odd CCKY forms and even KY forms are only defined in even dimensions and hence, this type of extension of the Poincare algebra is only possible for even dimensions \cite{Acik Ertem21}.

Now, we can define the brackets of this extension of Poincare algebra. In fact, the bracket of the extended Poincare algebra can again be defined as the CKY bracket. For the elements of $\Lambda^{odd}M$ such as a special odd CCKY $p$-form $\omega_1$ and a special odd CCKY $q$-form $\omega _2$, we have
\[
[\omega_1, \omega_2]_{CKY}=\frac{(-1)^p}{n-q+1}\omega_1\wedge\delta\omega_2+\frac{1}{n-p+1}\delta\omega_1\wedge\omega_2
\]
and the action of $\Lambda^{odd}M$ on $\Lambda^{even}M$ is defined as the $q$-form projection of the Clifford bracket
\[
([\delta\omega_1, \omega_2]_{Cl})_q=-\frac{4}{(n-p+1)!}\delta\omega_1\underset{\frac{p-1}{2}}{\wedge}\omega_2.
\]
The bracket for $\Lambda^{even}$ elements is written as the $(p+q-2)$-form projection of the Clifford bracket
\[
([\delta\omega_1, \delta\omega_2]_{Cl})_{p+q-2}=-\frac{(p+q)}{(n-p+1)(n-q+1)}\delta\omega_1\wedge\delta\omega_2
\]
From the coderivative of the SN bracket of $\omega_1$ and $\omega_2$ which can be found as
\[
\delta[\omega_1,\omega_2]_{CKY}=-\frac{(p+q)}{(n-p+1)(n-q+1)}\delta\omega_1\wedge\delta\omega_2
\]
and one can write the bracket for $\Lambda^{even}M$ as follows
\[
([\delta\omega_1, \delta\omega_2]_{Cl})_{p+q-2}=\delta[\omega_1,\omega_2]_{CKY}.
\]

Now, we can extend the Poincare superalgebra by defining the even part as the extended Poincare algebra. The odd part of the extended superalgebra will still be given by the space of Killing spinors $S$. The extended superalgebra is $e_0\oplus e_1=\Lambda^{odd}V\oplus\Lambda^{even}V\oplus S$ and the brackets are defined as follows. The even-even bracket
\[
[\,,\,]:e_0\times e_0\rightarrow e_0
\] can be defined as the brackets of the extended Poincare algebra. The even-odd bracket
\[
[\,,\,]:e_0\times e_1\rightarrow e_1
\]
is defined as the symmetry operators of Killing spinors
\begin{eqnarray}
[(\omega, -\frac{2}{(p-1)!(n-p+1)^2}\delta\omega\underset{\frac{p-1}{2}}{\wedge}), \psi]&=&L_{\omega}\psi\nonumber\\
&=&(i_{X^a}\omega).\nabla_{X_a}\psi+\frac{p}{2(n-p+1)}(\delta\omega)^{\xi}.\psi\nonumber\\
&=&(-(-1)^p\lambda p\omega+\frac{p}{2(m-p+1)}(\delta\omega)^{\xi}).\psi
\end{eqnarray}
where we have used the Killing spinor equation. The odd-odd bracket
\[
[\,,\,]:e_1\times e_1\rightarrow e_1
\]
is defined as
\[
[\psi_1,\psi_2]=\bigg((\psi_1\overline{\psi_2})_{p},-\frac{2}{(p-1)!(n-p+1)^2}\delta(\psi_1\overline{\psi_2})_{p}^{\xi}\underset{\frac{p-1}{2}}{\wedge}\bigg)
\]
where for a Killing spinor $\psi$, the odd degree spinor bilinear form $(\psi_1\overline{\psi_2})_{p}$ corresponds to a special odd KY $p$-form. 

For the extended Poincare superalgebra, Jacobi identities cannot be satisfied for all brackets. The even-even-even graded Jacobi identity is satisfied due to the properties of the SN bracket. However, the other Jacobi identities are not satisfied in general. So, it consists of a superalgebra structure but not a Lie superalgebra structure.

Poincare superalgebra is an important tool for understanding the relevant geometries that support supersymmetric theories and especially for the classification of supergravity backgrounds in various dimensions. It is known that the filtered subdeformations of the Poincare superalgebra can be obtained via the methods of Spencer cohomology and these subdeformations can be used in the classification of supergravity backgrounds in different dimensions. Extended Poincare superalgebras defined in this paper correspond to more general geometric structures on a background and possible extensions of the classification scheme for supergravity theories can be investigated by using these extended structures. For this to be done, possible subalgebras of the extended Poincare superalgebra which correspond to a Lie superalgebra and more general than the Poincare superalgebra can be investigated by satisfying the relevant Jacobi identities and the Spencer cohomology techniques can be apllied to these new superalgebras to obtain new classification results.

\section{Conclusion}

Isometries of a manifold generate symmetry algebras and they are related to special spinor fields called Killing spinors. Space of Killing spinors, their rotational representation and space of Killing spinors together constitute the Poincare superalgebra structure. Filtered subdeformations of the Poincare superlagebra is important in the classification of supergravity backgrounds in various dimensions. Besides the isometries, one can also construct higher degree form generalizations of them which are called hidden symmetries. We construct extended symmetry algebras to include the hidden symmetries and the extension of the Poincare superlagebra is also defined in terms of hidden symmetries and Killing spinors. The even part of the extended superalgebra consists of the direct sum of the space of special odd KY forms and the space of special even CCKY forms and the odd part is the space of Killing spinors. There is also a possible extension in even dimensions whose even part consists of the direct sum of the space of special odd CCKY forms and the space of special even KY forms. The definition of the extended Poincare superalgebra can give way to possible extensions of the classification scheme of supergravity backgrounds from the investigation of the filtered subdeformations of the extended superalgebra. To this aim, the conditions for satisfying a Lie superalgebra structure can be investigated and different information for physical theories can be extracted.

\begin{acknowledgments}

This study was supported by Scientific and Technological Research Council of T\"urkiye (T\"UB\.ITAK) under the Grant Number 123F261. The authors thank to T\"UB\.ITAK for their supports.

\end{acknowledgments}





\begin{thebibliography}{99}

\bibitem{Lichnerowicz} A. Lichnerowicz, Killing spinors, twistor spinors and Hijazi inequality, J. Geom. Phys. 5 (1988) 1.

\bibitem{Aleksy Cortes} D. V. Alekseevsky, V. Cortes, On pseudo-Riemannian manifolds with many Killing spinors, AIP Conf. Proc. 1093 (2009) 3. 

\bibitem{Klinker} F. Klinker, Supersymmetric Killing structures, Commun. Math. Phys. 255 (2005) 419.

\bibitem{Medeiros Hollands} P. de Medeiros, S. Hollands, Conformal symmetry superalgebras, Class. Quantum Grav. 30 (2013) 175016.

\bibitem{ACDV} D. V. Alekseevsky, V. Cortes, C. Devchand, A. Van Proeyen, Polyvector super-Poincare algebras, Commun. Math. Phys. 253 (2005) 385.

\bibitem{OFarrill Santi1} J. Figueroa-O'Farrill, A. Santi, Spencer cohomology and 11-dimensional supergravity, Commun. Math. Phys. 349 (2017) 627.

\bibitem{OFarrill Santi2} J. Figueroa-O'Farrill, A. Santi, On the algebraic structure of Killing superalgebras, Adv. Theor. Math. Phys. 21 (2017) 1115.

\bibitem{Yano} K. Yano, Some remarks on tensor fields and curvature, Ann. Math. 55 (1952) 328-347.

\bibitem{Semmelmann} U. Semmelmann, Conformal Killing forms on Riemannian manifolds, Math. Z. 245 (2003) 503.

\bibitem{Kastor Ray Traschen} D. Kastor, S. Ray, J. Traschen, Do Killing-Yano tensors form a Lie algebra?, Class. Quantum Grav. 24 (2007) 3759.

\bibitem{AEOV} \"O. A\c{c}{\i}k, \"U. Ertem, M. \"Onder, A. Ver\c{c}in, Basic gravitational currents and Killing-Yano forms, Gen. Relativ. Gravit. 42 (2010) 2543. 

\bibitem{Acik Ertem21} \"O. A\c{c}{\i}k, \"U. Ertem, Generalized symmetry superalgebras, J. Math. Phys. 62 (2021) 052303.

\bibitem{Ertem16} \"U. Ertem, Lie algebra of conformal Killing-Yano forms, Class. Quantum Grav. 22 (2016) 125033.


\end{thebibliography}
\end{document}